\documentclass[11pt]{article} 
\usepackage{rldmsubmit,palatino}
\usepackage{graphicx}
\usepackage[numbers]{natbib}
\usepackage{float}
\usepackage{amsmath}
\usepackage{hyperref}
\usepackage{subcaption}
\usepackage{cleveref}
\usepackage{sidecap} 

\title{Balancing Benefits and Risks: RL Approaches for Addiction-Aware Social Media Recommenders}

\author{
Luca Bolis, Stefano Livella, Sabrina Patania, Dimitri Ognibene \\
University of Milan-Bicocca\\
Milan, Italy\\
\texttt{sabrina.patania@unimib.it, dimitri.ognibene@unimib.it} \\
\And
Matteo Papini \\
Politecnico di Milano\\
Milan, Italy \\
\And
Kenji Morita \\
University of Tokyo \\
Tokyo, Japan \\
\And
}
\begin{document}

\maketitle

\begin{abstract}

Social media platforms provide valuable opportunities for users to gather information, interact with friends, and enjoy entertainment. However, their addictive potential poses significant challenges, including overuse and negative psychological or behavioral impacts \citep{alter2017irresistible, allcott2022digital, ognibene2023challenging}. This study explores strategies to mitigate compulsive social media usage while preserving its benefits and ensuring economic sustainability, focusing on recommenders that promote balanced usage.

We analyze user behaviors arising from intrinsic diversities and environmental interactions, offering insights for next-generation social media recommenders that prioritize well-being. Specifically, we examine the temporal predictability of overuse and addiction using measures available to recommenders, aiming to inform mechanisms that prevent addiction while avoiding user disengagement \citep{ognibene2019addiction}.

Building on RL-based computational frameworks for addiction modelling \citep{kato2023computational}, our study introduces:
\begin{itemize}
    \item A recommender system adapting to user preferences, introducing non-stationary and non-Markovian dynamics.
    \item Differentiated state representations for users and recommenders to capture nuanced interactions.
    \item Distinct usage conditions—light and heavy use—addressing RL’s limitations in distinguishing prolonged from healthy engagement.
    \item Complexity in overuse impacts, highlighting their role in user adaptation \citep{ognibene2019addiction}.
\end{itemize}

Simulations demonstrate how model-based (MB) and model-free (MF) decision-making interact with environmental dynamics to influence user behavior and addiction. Results reveal the significant role of recommender systems in shaping addiction tendencies or fostering healthier engagement. These findings support ethical, adaptive recommender design, advancing sustainable social media ecosystems \citep{zou2019reinforcement, agarwal2024system}.

\end{abstract}



\startmain 

\section{Introduction}

Social media platforms have transformed communication but raise concerns about compulsive use and addiction, driven by recommendation systems optimising engagement at the cost of well-being \citep{ognibene2023challenging, almourad2020defining}. Unlike pharmacological addiction, social media overuse stems from cognitive limitations and the inability to balance immediate rewards with long-term benefits \citep{ognibene2019addiction, kato2023computational}.

Reinforcement Learning (RL) frameworks capture these behaviours, modelling interactions between model-based (MB) and model-free (MF) decision-making. However, current models often neglect the adaptive, non-stationary influence of recommender systems on user engagement \citep{ognibene2023challenging}.

This study enhances RL addiction models by explicitly representing recommender systems, incorporating non-Markovian dynamics, differentiated state representations, and scenarios of light and heavy use. Simulations reveal interactions between MF and MB systems and recommender policies, offering strategies to mitigate addiction while promoting healthier engagement.

\section{Methods}
We define a multi-agent system consisting of a user and a recommender system operating within a shared environment of psycho-physical states.

\subsection{User modeling}
The user is represented by a dual RL system combining a Model-Free (MF) approach, implementing habitual behaviors using Q-learning to maximize cumulative rewards based on past experiences, and a Model-Based (MB) approach: Simulates planning and problem-solving using Q-Value Iteration, leveraging an internal representation of the environment for goal-directed decision-making. A parameter (Model Based Updates per Step) that determines how many updates of Q-Values are made at each iteration.

The two approaches were tested both individually and in combination.

\subsection{Recommender System modelling}
The recommender system in this study is implemented as a non-stationary multi-armed bandit (MAB) model, allowing it to dynamically adapt to user preferences through a balance of exploration and exploitation. Each arm of the bandit corresponds to a content recommendation, with rewards derived from user interactions—positive for acceptance and negative or neutral for rejection, reflecting evolving user behavior. This dynamic adaptation is crucial for simulating realistic interactions, as user preferences are often non-static. The non-stationary dynamics are modeled using an exponentially weighted average to prioritize recent interactions while discounting older data, enabling the recommender to adjust to temporal shifts in user behavior \citep{zou2019reinforcement, agarwal2024system}.

Moreover, distinct configurations differentiate short-term and prolonged social media usage. Short-term interactions yield moderate rewards with minimal penalties, while prolonged usage risks significant aftereffects, incorporating a feedback loop between user states and recommender policies. This approach captures the complexity of user-recommender interactions and underscores the potential for recommender systems to influence user behavior subtly but significantly over time.

\subsection{Environment design}

\begin{figure}[H]
    \centering
    \begin{subfigure}[t]{0.45\textwidth}
        \centering
        \includegraphics[width=\textwidth]{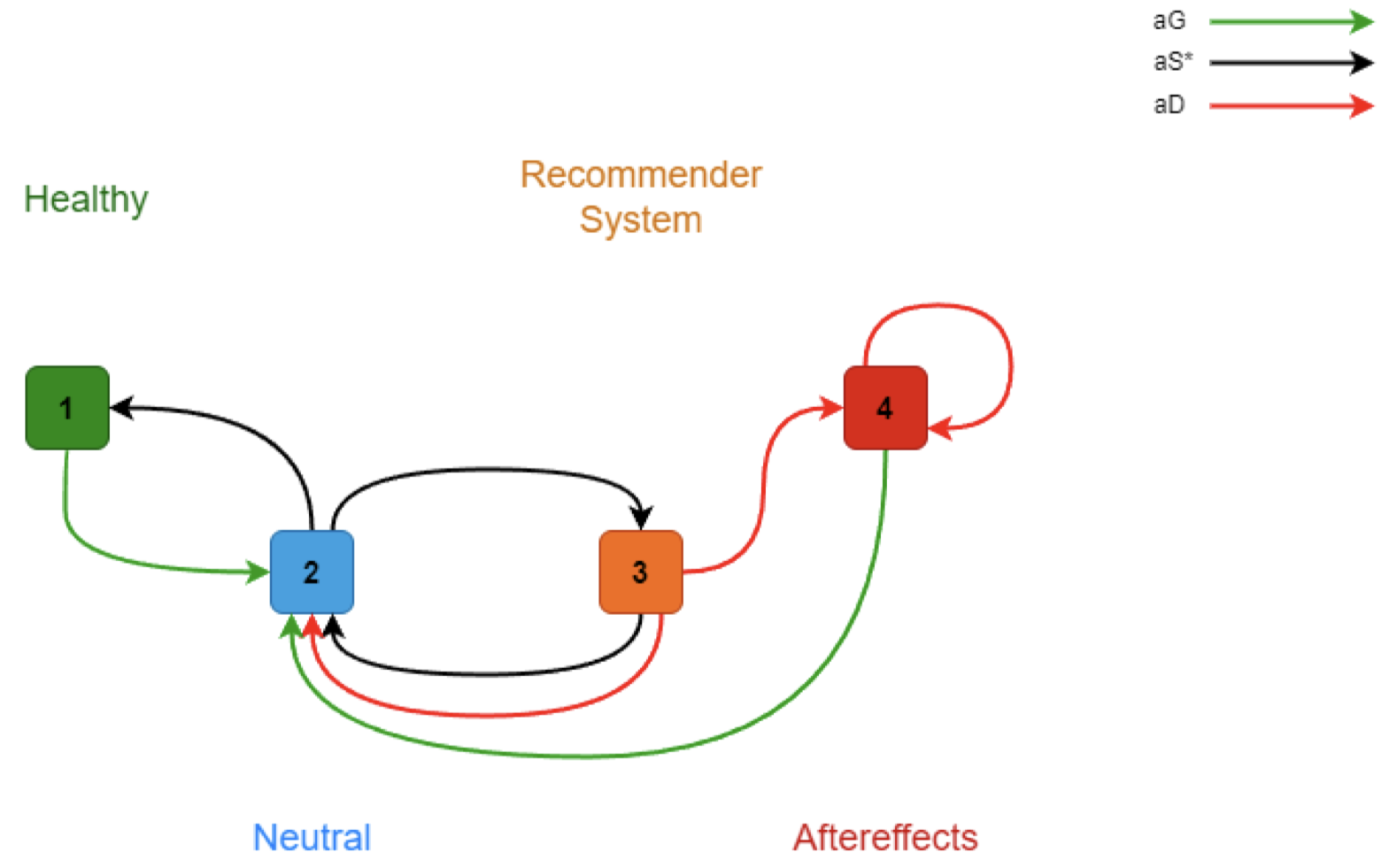}
        \label{fig:simple}
    \end{subfigure}
    \hfill
    \begin{subfigure}[t]{0.52\textwidth}
        \centering
        \includegraphics[width=\textwidth]{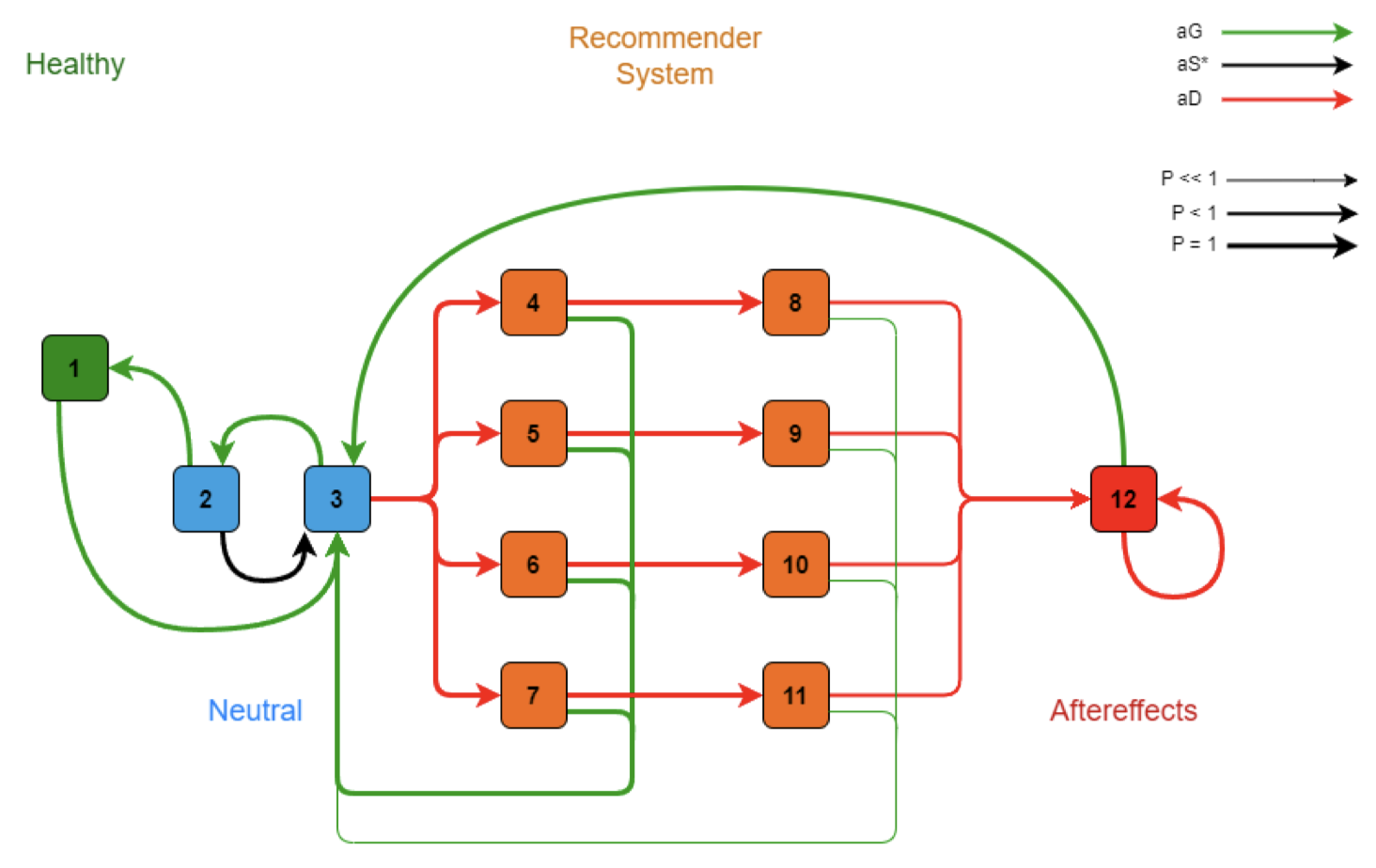}
        \label{fig:refined}
    \end{subfigure}
    \caption{Representation of states and actions for the simplified environment (left) and the advanced environment with refined recommender system (right).}
    \label{fig:env_comparison}
\end{figure}

The environment models user states (\texttt{Healthy}, \texttt{Neutral}, \texttt{Recommender Interaction}, and \texttt{Aftereffects}), actions, and transitions. Three levels of complexity were implemented:
\begin{enumerate}
    \item Simplified: a minimal structure for debugging and theoretical validation.
    \item Advanced: a detailed structure reflecting real-world variability.
    \item Refined Recommender System: differentiating short and prolonged social media usage to capture nuanced behavioral impacts (see Figure \ref{fig:env_comparison}).
\end{enumerate}


\section{Results}
To study the performance of our models, 900 simulations were performed
for each parameter combination and we considered the average of the individual tests. 
To analyze the simulation results, two charts were created, focusing on the key entities: the User and the Recommender System.

\begin{itemize}
    \item \textbf{Evolution of agents:} This graph aims to show how many users at a given iteration have developed a suboptimal policy and are, therefore, addicted. An agent that does not follow the optimal policy is defined as an addicted agent.
    \item \textbf{Q-values of the recommender system:} This graph exposes one of the key components of the recommendation system choice process, the Q-values. It shows which content (arm) is preferred by the agent; in particular, for each iteration we can see the amount of reward the recommender system expects to get if it were to present it.
\end{itemize}

\subsection{ Key differences in model parameterization:}
\begin{figure}[H]
    \centering
    \begin{subfigure}[t]{0.48\textwidth}
        \centering
        \includegraphics[width=\textwidth]{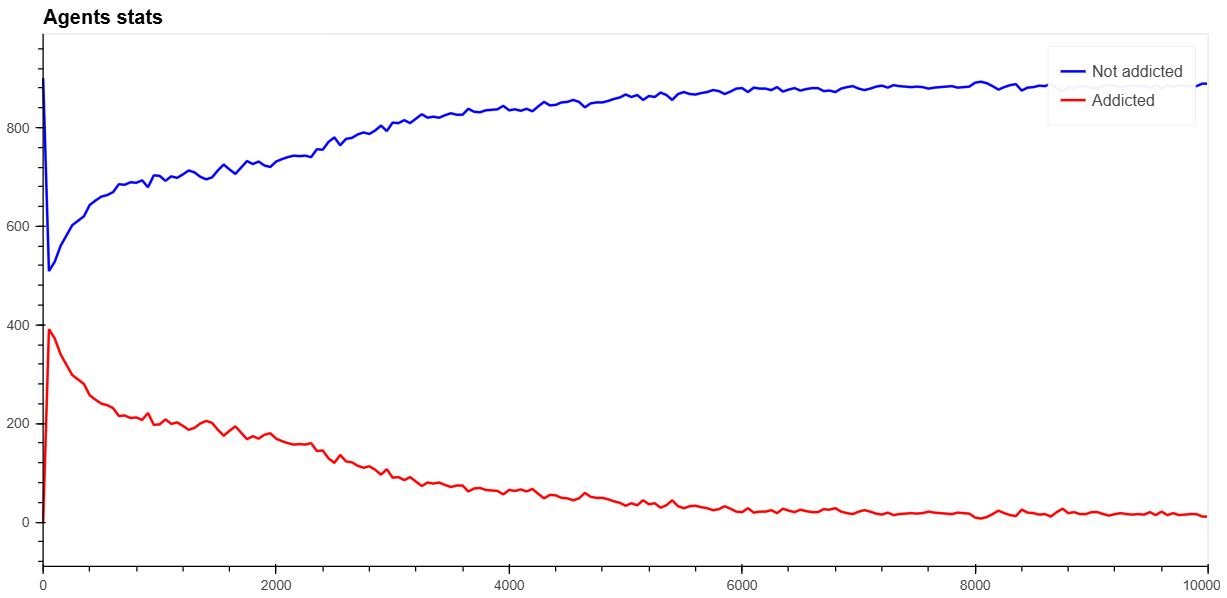}
        \caption{$\beta$=0.25}
        \label{subfig1labela}
    \end{subfigure}
    \hfill
    \begin{subfigure}[t]{0.48\textwidth}
        \centering
        \includegraphics[width=\textwidth]{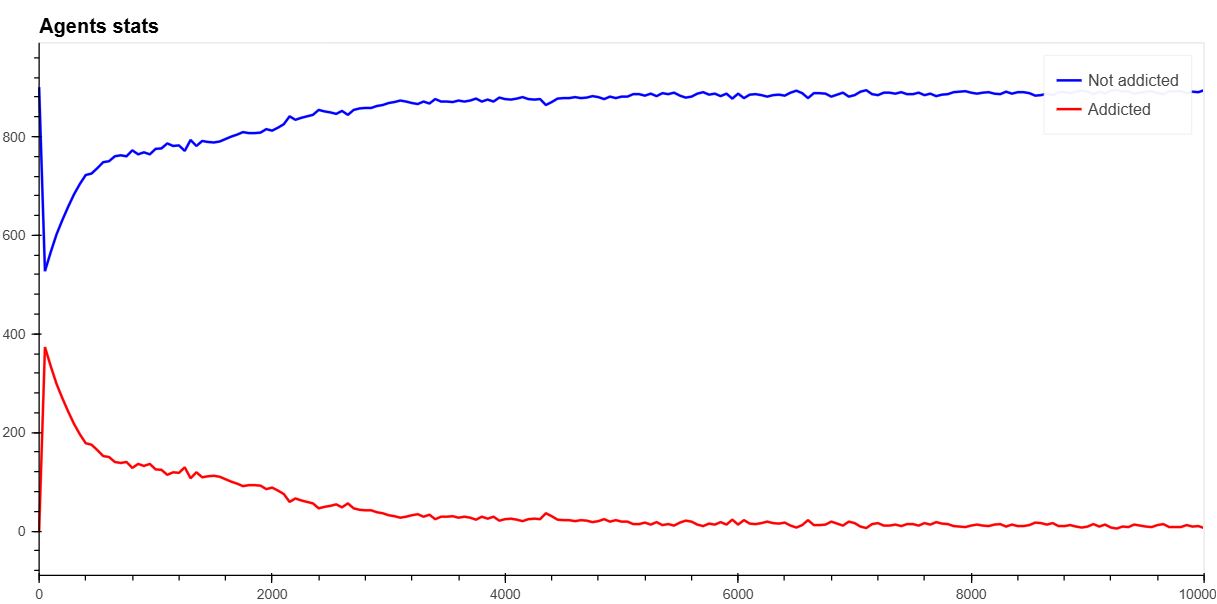}
        \caption{$\beta$=0.75}
        \label{subfig1labelb}
    \end{subfigure}
    \caption{Number of agents who identified the optimal policy in the environment with the refinement of the recommender system.}
    \label{fig:fig1}
\end{figure}

A Dual RL model system comprises two components, Model-Free (MF) and Model-Based
(MB), that cooperate to identify the optimal policy. The impact of MF and MB on the
expected rewards is determined by a parameter $\beta$. The Q-Values for the Dual RL model (\(Q_M\)) are calculated as:
\[
Q_M(s, a) = \beta Q_{MB}(s, a) + (1 - \beta) Q_{MF}(s, a).
\]
In the simplified, advanced and refined environment, increasing the value of $\beta$ results in a higher number of non-addictive agents as you can see in the \autoref{fig:fig1}, since it's fully capable of building the entire model of the environment.

\textbf{MBUS} (Model Based Updates per Step) is a parameter that determines how many
updates of Q-Values are made at each iteration. A higher value increases the time
required to complete the operation but results in more reliable estimates and improvement in performance. 

\begin{figure}[H]
    \centering
    \begin{subfigure}[t]{0.48\textwidth}
        \centering
        \includegraphics[width=\textwidth]{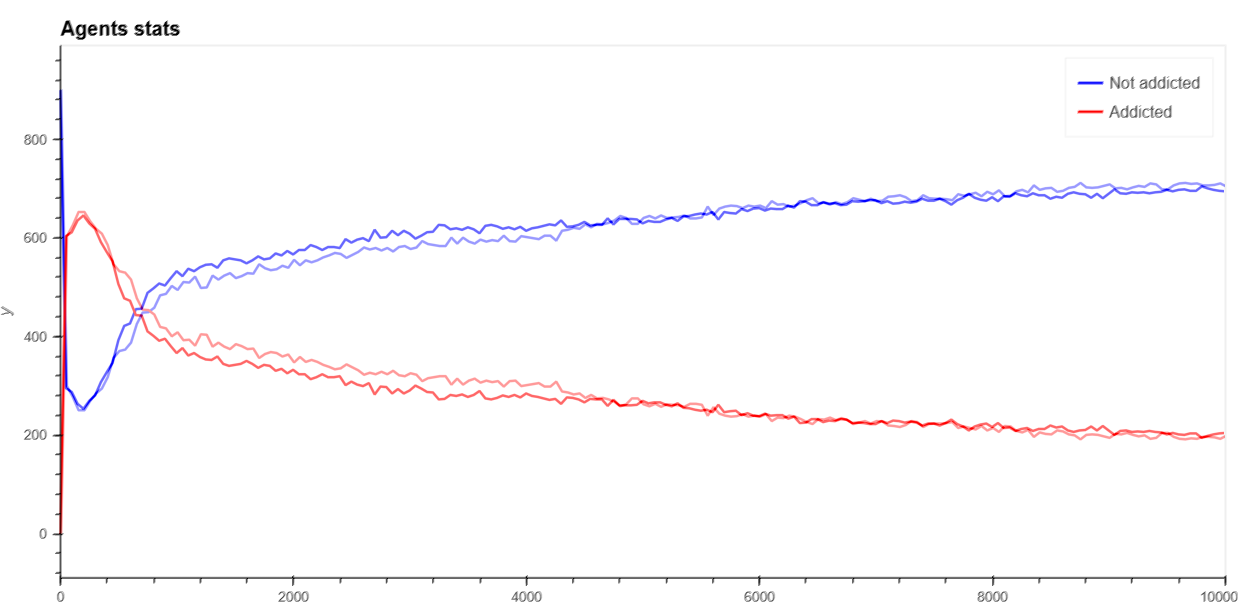}
        \caption{MBUS performance comparison in the misrepresented environment}
        \label{fig:fig21}
    \end{subfigure}
    \hfill
    \begin{subfigure}[t]{0.48\textwidth}
        \centering
        \includegraphics[width=\textwidth]{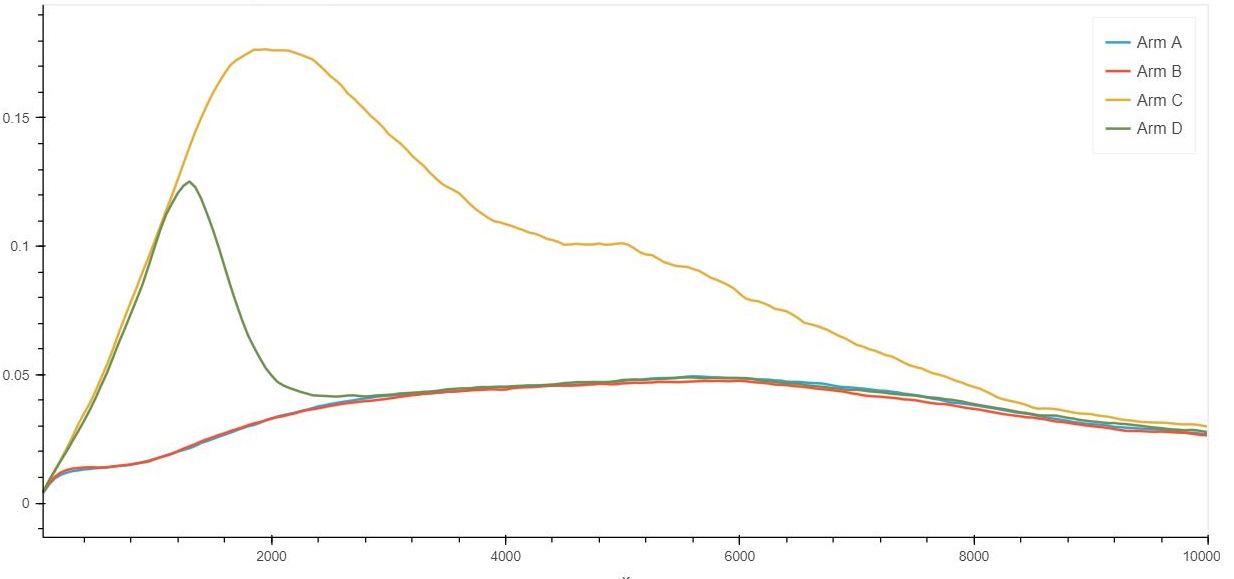}
        \caption{Q-values of the recommender system in the refined environment}
        \label{fig:fig22}
    \end{subfigure}
    \caption{(Left) MBUS performance comparison in the misrepresented environment. (Right) Q-values of the recommender system in the refined environment.}
    \label{fig:fig_combined}
\end{figure}

\autoref{fig:fig21} shows the comparison between two simulations performed with the same value of $\beta$ ($\beta$ = 0.5) and different MBUS values, 1 in blue and 50 in a lighter shade of blue.

We can see that the curve for agents that identified the optimal policy grows faster at the
beginning with MBUS = 50, an indication that on average users need fewer steps to find
the best policy.

In the refined environment, the recommender system has to discover which one of the four arms gives the user the best rewards.
As you can see in the \autoref{fig:fig22} produced in the analysis of the results, it is fully capable of doing so and, therefore, influence the user towards an addictive behavior.

\begin{figure}[H]
    \centering
    \begin{subfigure}[t]{0.48\textwidth}
        \centering
        \includegraphics[width=\textwidth]{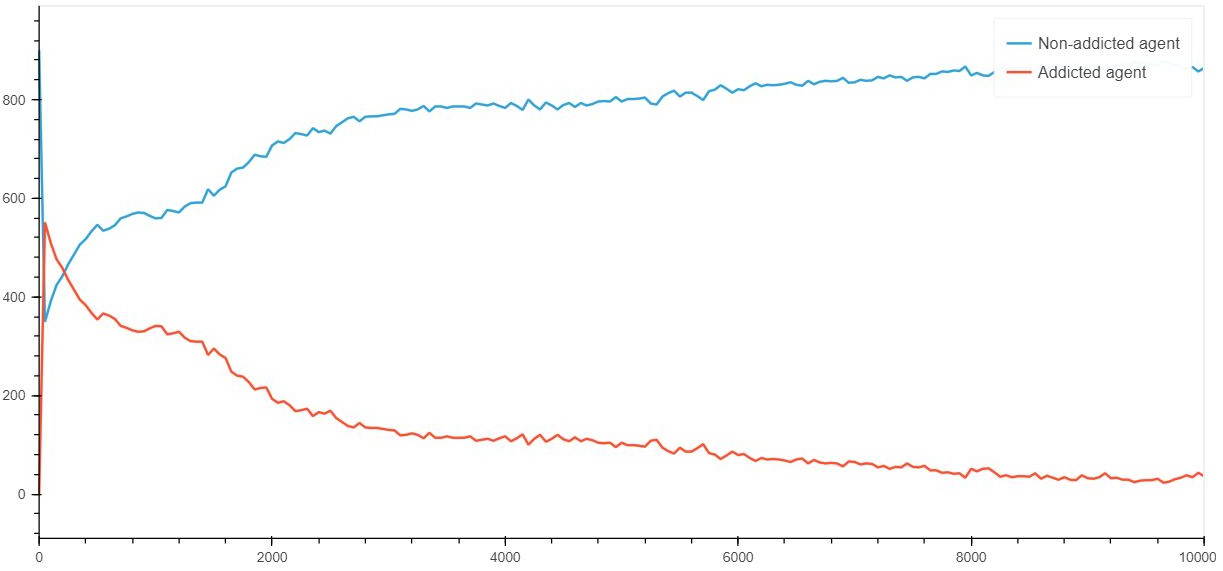}
        \caption{Number of agents identifying the optimal policy with $\beta$=0}
        \label{fig:fig31}
    \end{subfigure}
    \hfill
    \begin{subfigure}[t]{0.48\textwidth}
        \centering
        \includegraphics[width=\textwidth]{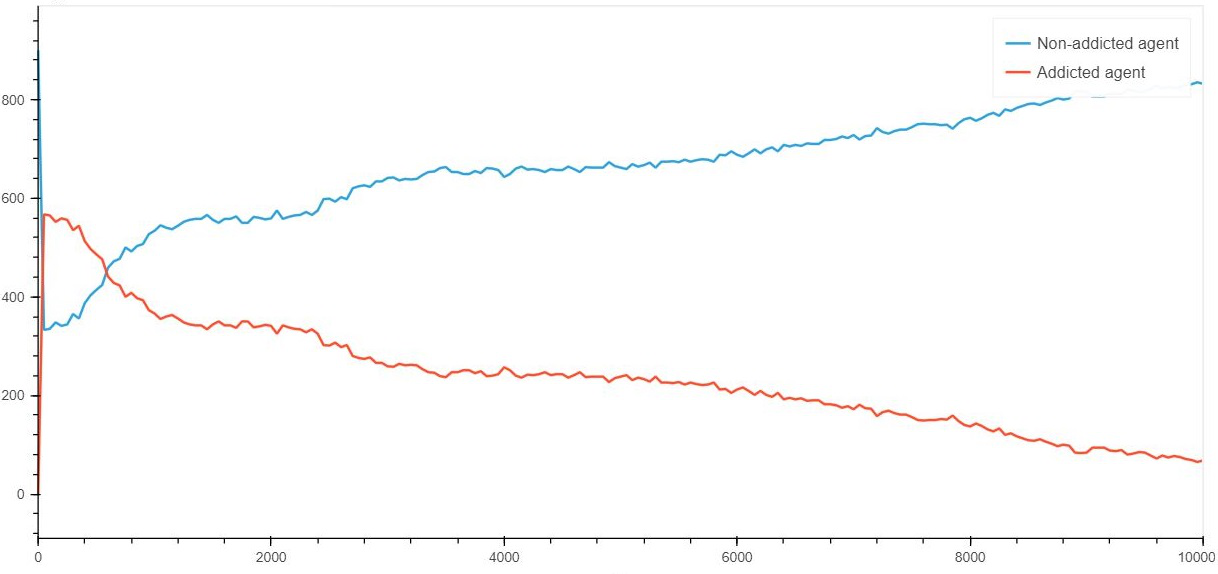}
        \caption{Number of agents identifying the optimal policy with $\beta$=0.5}
        \label{fig:fig32}
    \end{subfigure}
    \caption{Number of agents who identified the optimal policy in scenarios with misrepresentation of the environment, for $\beta$=0 and $\beta$=0.5.}
    \label{fig:fig3_combined}
\end{figure}

\subsection{Misrepresentation of the Environment:}
The results seen so far demonstrate how MB, due to its ability to build an
accurate internal representation of the world, guarantees the best results. In
order to create a more realistic context, transition
functions related to the Healthy state were modified, leaving the probability
of reaching it equal to 0 throughout the simulation, thus
making the Healthy state invisible to the user.
The \autoref{fig:fig3_combined} shows the results for this environment with $\beta$=0, and for $\beta=0.5$, in both cases, there is a significantly higher number of addicted agents.


\section{Conclusion and future directions}


The chosen approach produced results consistent with real-world observations and prior studies \cite{ognibene2019addiction}, validating the dual RL model system for decision-making. Simulation graphs reveal that users with perfect knowledge of their environment maintain healthy behaviours and avoid addiction, whereas those with imperfect knowledge—resembling real-world conditions—become more dependent on social media. The subtle harms of excessive social media use often go unnoticed until they significantly impact well-being.

Our multi-armed recommender system highlights how user preferences can be effectively identified but also exploited, fostering dependence.

The project’s core components offer several directions for future work:
\begin{itemize}
\item \textbf{Decision-making mechanisms:} Explore alternatives like Active Inference, which minimises uncertainty rather than maximising rewards \citep{friston2017active}.
\item \textbf{Environment modelling:} Increase states and actions to simulate more realistic scenarios.
\item \textbf{Recommender systems:} Incorporate user-specific attributes (e.g., age, nationality, social role) to enhance profiling and personalisation.
\end{itemize}

\bibliographystyle{plain}
\bibliography{rldm}

\end{document}